\documentclass[aip,amsmath,amssymb,reprint]{revtex4-1}
\usepackage{mathptmx}
\usepackage[T1]{fontenc}
\usepackage[utf8]{luainputenc}
\usepackage{color}
\usepackage{mathrsfs}
\usepackage{amsmath}
\usepackage{graphicx}

\makeatletter
\usepackage{dcolumn}
\usepackage{bm}

\usepackage{etoolbox}
\usepackage{dcolumn}
\usepackage{bm}

\usepackage{etoolbox}

\def\@email#1#2{%
 \endgroup
 \patchcmd{\titleblock@produce}
  {\frontmatter@RRAPformat}
  {\frontmatter@RRAPformat{\produce@RRAP{*#1\href{mailto:#2}{#2}}}\frontmatter@RRAPformat}
  {}{}
}%

\makeatother

\begin{document}
\preprint{AIP/123-QED}
\title{Run-and-tumble particles in a harmonic external field:\\
 constructing the solution of the Fokker-Planck equation from the
equations of motion }
\author{Aoran Sun}
\email{Corresponding author. sunaoran16@mails.ucas.ac.cn}

\affiliation{Beijing National Laboratory for Condensed Matter Physics, Institute
of Physics, Chinese Academy of Sciences, Beijing 100190, China}
\affiliation{Zhejiang Key Laboratory of Soft Matter Biomedical Materials, Wenzhou
Institute, University of Chinese Academy of Sciences, Wenzhou, Zhejiang
325000, China}
\author{Fangfu Ye}
\email{Corresponding author. fye@iphy.ac.cn}

\affiliation{Beijing National Laboratory for Condensed Matter Physics, Institute
of Physics, Chinese Academy of Sciences, Beijing 100190, China}
\affiliation{Zhejiang Key Laboratory of Soft Matter Biomedical Materials, Wenzhou
Institute, University of Chinese Academy of Sciences, Wenzhou, Zhejiang
325000, China}
\author{Jeff Z. Y. Chen}
\email{Corresponding author. jeffchen@uwaterloo.ca}

\affiliation{Zhejiang Key Laboratory of Soft Matter Biomedical Materials, Wenzhou
Institute, University of Chinese Academy of Sciences, Wenzhou, Zhejiang
325000, China}
\affiliation{Department of Physics and Astronomy, University of Waterloo, Waterloo,
Ontario N2L 3G1, Canada}
\date{\today}
\begin{abstract}
A method for integrating the Langevin equations of motion for run-and-tumble
particles is introduced to compute the joint distribution of their
position and active velocity, with and without a harmonic external
force field. The key step is the construction of the moments of the
distribution and the correlation functions, conditional on the particles
being in a specific state at the final time. For one-dimensional run-and-tumble
particles in a harmonic external potential, the exact steady-state
distributions for the tumble state, as well as for the run state conditional
on a specified velocity direction, are obtained. This enables us to
demonstrate their equivalence to the exact solutions of the corresponding
Fokker--Planck equations. 
\end{abstract}
\maketitle

\section{Introduction}

Active particles differentiate themselves from passive particles by
their characteristic ability to dissipate energy from their environment
to fuel their motions \citep{2013_HydrodynamicsofSoftActiveMatter,2016_ActiveParticlesinComplexandCrowdedEnvironments,2022_IrreversibilityandBiasedEnsemblesinActiveMatterInsightsfromStochasticThermodynamics,2026_ColloquiumWhatDoWeMeanbyactiveMatter}.
Well-known examples of active systems include molecular motors \citep{2006_ActiveGelsDynamicsofPatterningandSelfOrganization,2007_NonequilibriumMechanicsofActiveCytoskeletalNetworks},
living cells \citep{2004_E.ColiinMotion,2008_OutofEquilibriumMicrorheologyinsideLivingCells},
energetically excited granular materials \citep{2017_NoiseandDiffusionofaVibratedSelfPropelledGranularParticle},
active gels \citep{2010_TheMechanicsandStatisticsofActiveMatter,2019_SignaturesofMotorSusceptibilitytoForcesintheDynamicsofaTracerParticleinanActiveGel},
or even on a much larger scale, multicellular animals \citep{2004_AModeloftheFormationofFishSchoolsandMigrationsofFish,2005_HydrodynamicsandPhasesofFlocks,2014_FlockingataDistanceinActiveGranularMatter,Du_2021}.
Many have taken a step further by fabricating artificial active matter,
designed for special functionalities \citep{2014_GravitaxisofAsymmetricSelfPropelledColloidalParticles,2016_AcousticTrappingofActiveMatter,2016_ActiveParticlesinComplexandCrowdedEnvironments}.
This class of systems form a major research area in recent years,
because the physics that governs them underlies many real systems
in nature.

Active particles are theoretically a paradigmatic platform for nonequilibrium
and non-Boltzmann statistics, a topic that remains underdeveloped
\citep{2014_GravitaxisofAsymmetricSelfPropelledColloidalParticles,2015_ActiveBrownianParticlesandRunandTumbleParticlesaComparativeStudy,2016_AcousticTrappingofActiveMatter}.
The behavior of a single active particle in free space can be non-trivial
or even counter-intuitive. Two types of minimal models are often used:
the active Brownian particle (ABP) and the run-and-tumble particle
(RTP). To some extent, the ABP can successfully model many known microswimmers
\citep{2012_ActiveBrownianParticlesfromIndividualtoCollectiveStochasticDynamics,2016_ActiveParticlesinComplexandCrowdedEnvironments,2025_ThermodynamicsofActiveMatterTrackingDissipationacrossScales,2026_ColloquiumWhatDoWeMeanbyactiveMatter},
while the RTP model was inspired by the motion of certain types of
bacteria, such as Escherichia coli \citep{2004_E.ColiinMotion,2022_WrappedUptheMotilityofPolarlyFlagellatedBacteria}.
In both cases, the magnitude of the particle's velocity is assumed
fixed, and its direction changes stochastically over time. An ABP
constantly changes the velocity direction. By contrast, an RTP maintains
its velocity direction for an exponentially distributed time duration,
within the "run" state, then enters a "tumble" state at which
it randomly selects a new velocity direction, followed by yet another
run state. The RTP model exhibits rich dynamic features that have
only begun to be established, mostly by computer simulations, in recent
years \citep{2015_ActiveBrownianParticlesandRunandTumbleParticlesaComparativeStudy,2020_RunandTumbleParticlesinTwoDimensionsMarginalPositionDistributions,2020_UniversalSurvivalProbabilityforadDimensionalRunandTumbleParticle}.
The present paper focuses on the analytical treatment of the dynamics
of RTPs.

For a free single RTP in the absence of an external field, the time
evolution of the probability density function of position has been
analytically studied using Laplace transformation in time and Fourier
transformation in space \citep{2013_AveragedRunandTumbleWalks}. A
key property for the RTP is the first-passage time, the time an RTP
takes to reach a target position for the first time, given its first
position. This quantity captures, for example, the RTP's ability to
search and forage, and has therefore been studied in various settings
\citep{2019_NoncrossingRunandTumbleParticlesonaLine,2020_UniversalSurvivalProbabilityforadDimensionalRunandTumbleParticle,2025_OptimalFirstPassageTimesofActiveBrownianParticlesunderStochasticResetting,2025_Encounterbasedmodelofarunandtumbleparticlewithstochasticresetting}.
Other physically interesting quantities include the time to reach
the maximum distance up to a given time, as well as the time it takes
to return to its starting position \citep{2019_GeneralisedArcsineLawsforRunandTumbleParticleinOneDimension}.
The problem of a single ABP appears to be harder to analyze, with
all known analytic results for the time evolution of its position
probability density function expressible only as complicated series
expansions \citep{2015_SmoluchowskiDiffusionEquationforActiveBrownianSwimmers,2016_IntermediateScatteringFunctionofanAnisotropicActiveBrownianParticle}.

Active particles moving in an external potential represent another
important class of problems \citep{2015_RunandTumbleDynamicsofSelfPropelledParticlesinConfinement,2020_VelocityandDiffusionConstantofanActiveParticleinaOneDimensionalForceField,2024_ConfinedRunandTumbleParticleswithNonMarkovianTumblingStatistics,2025_IntegralEquationFormulationofRunandTumbleParticlesinaHarmonicTrapSpecialStatusofaSysteminTwoDimensions,2026_ExactStationaryStateofadDimensionalRunandTumbleParticleinaHarmonicPotential}.
It is expected that at long times, the active particle can eventually
settle into a non-Boltzmann, non-equilibrium steady state \citep{2022_ExactPositionDistributionofaHarmonicallyConfinedRunandTumbleParticleinTwoDimensions,2022_PositingtheProblemofStationaryDistributionsofActiveParticlesAsThirdOrderDifferentialEquation,2023_RunandTumbleOscillatorMomentAnalysisofStationaryDistributions,2024_ActiveOscillatorRecurrenceRelationApproach}.
A special case is the harmonic potential, which models the optical
\citep{1997_OpticalTrappingandManipulationofNeutralParticlesUsingLasers}
or acoustic \citep{2016_AcousticTrappingofActiveMatter} tweezers
commonly used in experiments. For this potential, the steady-state
position distribution function for an RTP in one-dimensional (1D)
\citep{2015_PressureIsNotaStateFunctionforGenericActivefluids,2019_RunandTumbleParticleinOneDimensionalConfiningPotentialsSteadyStateRelaxationandFirstPassageProperties}
or two-dimensional (2D) space \citep{2022_PositingtheProblemofStationaryDistributionsofActiveParticlesAsThirdOrderDifferentialEquation,2023_RunandTumbleOscillatorMomentAnalysisofStationaryDistributions,2024_ActiveOscillatorRecurrenceRelationApproach,2026_ExactStationaryStateofadDimensionalRunandTumbleParticleinaHarmonicPotential}
has been found exactly; in three-dimensional (3D) space, beyond its
moments \citep{2022_PositingtheProblemofStationaryDistributionsofActiveParticlesAsThirdOrderDifferentialEquation,2023_RunandTumbleOscillatorMomentAnalysisofStationaryDistributions,2024_ActiveOscillatorRecurrenceRelationApproach},
the positional distribution is analytically expressible in integral
form \citep{2026_ExactStationaryStateofadDimensionalRunandTumbleParticleinaHarmonicPotential}.
The steady-state distribution function of an ABP in an external field
is harder to find and is only expressed in terms of series expansions
\citep{2020_SteadyStateofanActiveBrownianParticleinaTwoDimensionalHarmonicTrap,2022_AnalyticSolutionofanActiveBrownianParticleinaHarmonicWell};
the moments of this function have been expressed recursively as well
\citep{2024_ActiveOscillatorRecurrenceRelationApproach}.

Two commonly used theoretical frameworks for understanding the dynamics
are the Fokker-Planck equations and the Langevin equations. The former
deals with the time $t$ evolution of the joint probability density
function $f$ of the position $r$ and active velocity $u$, and is
usually expressed as a partial differential equation. For the motion
of RTPs in no external field\citep{2019_RunandTumbleParticleinOneDimensionalConfiningPotentialsSteadyStateRelaxationandFirstPassageProperties,2022_ExactPositionDistributionofaHarmonicallyConfinedRunandTumbleParticleinTwoDimensions,2022_PositingtheProblemofStationaryDistributionsofActiveParticlesAsThirdOrderDifferentialEquation}
the equation is written as: 
\begin{equation}
\frac{\partial f}{\partial t}+\nabla\left(uf\right)=\gamma\left[-f+\frac{1}{\left|\Omega\right|}\int_{\Omega}f\left(r,u,t\right)du\right]
\end{equation}
where $\gamma$ is the tumble rate.The second term in the square brackets
is the velocity-averaged $f$ over the velocity domain $\Omega$.
In continuous active velocity distribution, $\Omega$ is the complete
solid angle, and in discontinuous distribution where the integral
itself is rewritten as a sum, $\Omega$ is the total number of directions. 

The Langevin equations mainly concern the trajectories of the involved
particles. Whether in the form of a random force or directly as a
coupling to the velocity of the particles, the stochastic process
is expressed by an additional term in the equation of motion. One
then either samples the trajectories directly in a computer simulation,
for which the stochastic term is drawn from a probability distribution
function, or takes the route of determining the overall averaged properties
of the system, by analytically incorporating the stochastic probability
into the averaged physical quantities. For the current problem, the
Langevin dynamic equation is introduced in Sect. \ref{subsec:Langevin-equation}.

A useful treatment following the latter approach is the Kac method
\citep{1974_AStochasticModelRelatedtotheTelegraphersEquation}, which
directly integrates the Langevin equation to calculate the moments
of the position without going through the Fokker--Planck equation.
The positional probability distribution can then be deduced from its
moments. The original solution was limited to a 1D model, essentially
corresponding to a 1D RTP without an external potential. Extending
the method to more general RTP problems, it has been successfully
applied to RTPs in arbitrary dimensions and to 2D ABP problems \citep{2025_ExactMomentsforaRunandTumbleParticlewithaFiniteTumbleTimeinaHarmonicTrap},
both with and without a harmonic external force field. It also enables
the treatment of more detailed and complicated models, such as RTPs
with a finite tumble time between runs \citep{2025_ExactMomentsforaRunandTumbleParticlewithaFiniteTumbleTimeinaHarmonicTrap}
and two-dimensional active Brownian particle models \citep{2025_MomentAnalysisofTwoDimensionalActiveBrownianRunandTumbleParticles};
see also \citep{2025_TheKacMethodanAlternativetotheFokkerPlanckEquationsfortheMotionofActiveParticles}.

However, so far the original Kac method has a fundamental limitation:
it only yields the positional distribution of the particles. In contrast,
the Fokker--Planck formalism naturally provides the full joint distribution
of both position and the active velocity of the particle. The result
obtained from the Kac method can therefore be viewed as the marginal
distribution of the full Fokker--Planck solution, at a level of being
integrated over the and active velocities. As a consequence, the approach
discards additional dynamical information and makes it impossible
to directly verify the result by substitution of the marginal distribution
into the Fokker--Planck equation.

\begin{figure}
\centering

\includegraphics[width=0.9\linewidth]{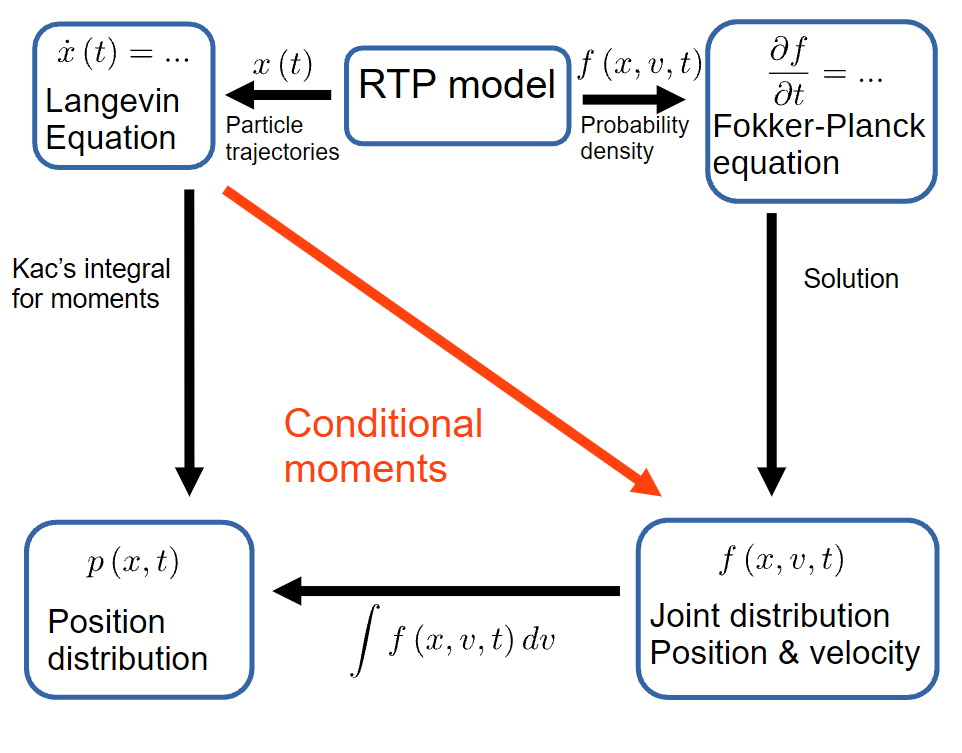} \caption{Schematic presentation of the goal of this work. In a theoretical
study of the RTP model, one may proceed along the right-hand route
in the figure by considering the evolution of the joint probability
density of position and active velocity, appears in the Fokker--Planck
equation. Alternatively, one may follow the left-hand route by considering
individual trajectories following the Langevin equations. Previously,
following the idea of Kac, the Langevin equations were used to compute
the moments of position, which ultimately lead to the time-dependent
positional distribution; in such an approach, information on the active
velocity is not retained. The main contribution of the present study
is the red route. The moments, conditional on the active velocity
at the final time, is computed directly from the Langevin equation.
From them the conditional distribution is obtained, and its equivalence
to the joint distribution (i.e., the solution of the Fokker--Planck
equation) is then demonstrated.}
\label{fig:ad} 
\end{figure}

This work aims to build a general theoretical framework for the joint
distributions, extending the methods of Kac. Building on the new findings,
a justification is made to link the results from the Langevin treatment
directly to solutions of the Fokker--Planck equation, which can only
be achieved when the active velocity is explicitly considered. The
idea is to consider the conditional distribution---i.e., the distribution
conditional on the particle being in a specific state at the final
time (e.g., run or tumble) and having its active velocity pointing
in a given direction---directly from the Langevin equation. This
is done by computing the conditional correlation functions, integrating
them to obtain the conditional moments, and finally reconstructing
the conditional distributions from these moments. A schematic presentation
of the idea is presented in Fig. \ref{fig:ad}, and a more detailed
outline of the method is provided in Sect. \ref{sec:Conditioned-moments}.

For active particles in harmonic traps, the derivation of moments
of the marginal distribution functions was developed in a series of
papers by Frydel \citep{2022_PositingtheProblemofStationaryDistributionsofActiveParticlesAsThirdOrderDifferentialEquation,2023_RunandTumbleOscillatorMomentAnalysisofStationaryDistributions,2024_ActiveOscillatorRecurrenceRelationApproach}.
That approach started from the Fokker--Planck equation itself, without
direct reference to the Langevin equation. In contrast, we extend
Kac's original idea to deal with active particles based on a stochastic
treatment of the Langevin equations. These are two distinct starting
points: ours follows the red path in Fig.~\ref{fig:ad}, whereas
Frydel's follows the vertical path on the right (though he obtained
the marginal distribution only).

For the specific cases of 1D RTPs with zero and finite tumble times,
we then proceed to Sects. \ref{sec:Run-and-tumble-particle} and \ref{sec:1D-two-state},
respectively. As an illustration of the general framework developed
in Sect. \ref{sec:Conditioned-moments}, the conditional probability
distribution functions for RTPs in an external harmonic force field
are explicitly calculated and expressed. In the steady-state limit,
these results are shown to be equivalent to the exact solutions of
the corresponding Fokker--Planck equations.

\section{Method Outline }

\label{sec:Conditioned-moments}

\subsection{Langevin equation}

\label{subsec:Langevin-equation}

The basic assumption here is that the motion of a RTP, possibly in
a harmonic potential, is governed by the generalized Langevin equation.
Regardless of the actual dimension of the model, only one coordinate
component is considered. The projected Langevin equation for this
coordinate component is written as: 
\begin{equation}
\dot{x}(t)=-bx(t)+F\bigl(t,K(t)\bigr)\label{eq:equation-of-motion}
\end{equation}
where $b$ is the strength of the harmonic potential, and can be set
to zero for a seamless transition to RTP in free space. The additional
term $F$ is the projection of the so-called active velocity, a stochastic
contribution that depends on time and on the internal state $K$.
The process $K(t)$ follows a Markov process with possibly $m$ states.
For an arbitrary state $A$ among these $m$ states, the magnitude
of the active velocity is $v_{A}$, and the transition rate from state
$A$ to state $B$ is $\gamma_{BA}$. For convenience, we define $\gamma_{A}=\sum_{B\neq A}\gamma_{BA}$
and introduce a matrix $\Lambda$ whose elements are given by $\Lambda_{AA}=-\gamma_{A}$
and $\Lambda_{AB}=\gamma_{AB}$ for $B\neq A$. In addition, it is
assumed that the Markov process $K$ is at equilibrium; therefore,
for any state $A$, the probability $\mathbb{P}(K(t)=A)$ is time-independent
and will be denoted by $\mathbb{P}(A)$.

While it is possible to generalize, in this work the case $m=2$ is
considered, consisting of a run state $\mathrm{R}$ and a tumble state
$\mathrm{T}$. In the tumble state, $v_{T}=0$. For simplicity, we
denote $v_{R}=v$, $\gamma_{RT}=\gamma_{T}$, and $\gamma_{TR}=\gamma_{R}$.
The transition matrix $\Lambda$ then takes the form 
\begin{equation}
\varLambda=\left(\begin{array}{cc}
-\gamma_{R} & \gamma_{T}\\
\gamma_{R} & -\gamma_{T}
\end{array}\right).
\end{equation}
Furthermore, at the beginning of each $\mathrm{R}$ state, the active
velocity is chosen randomly and independently, with fixed magnitude
$v$ but random orientation. Consequently, $F$ is uncorrelated between
different run states. In one dimension, the possible values of $F$
in the $\mathrm{R}$ states are ${\pm v}$.

\subsection{Moments of positional distribution }

\label{subsec:The-Kac-method}

A formal solution to Eq. \ref{eq:equation-of-motion}, the equation
of motion, is 
\begin{align}
x(t)=\int^{t}_{0}F\bigl(s,K(s)\bigr)e^{-b(t-s)}ds\label{eq:SolutionGeneralODE}
\end{align}
For simplicity, the initial condition $x(0)=0$ is assumed throughout
the remainder of this work.

Next, we calculate the moments of the time-dependent positional distribution:
\begin{align}
\left\langle x\left(t\right)^{l}\right\rangle = & ~l!\int_{0\leqslant t_{1}\leqslant...\leqslant t_{l}\leqslant t}\left(\prod^{l}_{i=1}dt_{i}\right)e^{-b\left(lt-\sum_{k}t_{k}\right)}\left\langle \prod^{l}_{k=1}F\left(t_{k}\right)\right\rangle .\label{eq:MomentsGeneral}
\end{align}
Here, for simplicity, the notation $F(t)=F\bigl(t,K(t)\bigr)$ is
used. The expectation $\langle...\rangle$ is taken over all realizations
of $F$. The symmetry properties of the integrand have been used to
reorder the time variables.

Although Eq. \eqref{eq:MomentsGeneral} follows directly from Eq.
\eqref{eq:SolutionGeneralODE}, relatively few studies, to the best
of our knowledge, have proceeded along this path to calculate moments
of a general order $l$. A notable exception is the work by Kac, who
studied the special case with $b=0$ and $F$ given by a dichotomous
random process \citep{1974_AStochasticModelRelatedtotheTelegraphersEquation}.
In the language of active matter, this corresponds to a 1D RTP without
interactions or external potential. He successfully obtained the Laplace
transform of $\langle x(t)^{l}\rangle$ for all positive integers
$l$.

To proceed with the general calculation, finding the correlation function
$\left\langle \prod_{k}F(t_{k})\right\rangle $ becomes essential.
For the RTP model, we adopt a case-by-case approach, evaluating the
contribution from each case and then summing over all cases. More
formally, we apply the law of total expectation \citep{2018_StochasticProcesses}:
\begin{equation}
\left\langle \cdot\right\rangle =\sum_{\mathrm{diag}}\left\langle \cdot\mid\mathrm{diag}\right\rangle \mathbb{P}(\mathrm{diag}),
\end{equation}
where $\mathrm{diag}$ denotes the different cases under consideration
and $\mathbb{P}(\mathrm{diag})$ is the probability of each case.
The classification of cases is based on the following two criteria:
(A) the state of the particle at each time $t_{k}$ appearing in the
correlation function; in practice, realizations with different values
of $K(t_{k})$ for any $k$ belong to different cases, whereas for
times $\tau$ not explicitly appearing in the correlation function
(i.e., $\tau\neq t_{k}$ for all $k$), realizations with different
$K(\tau)$ may belong to the same case. (B) the presence or absence
of state transitions between specified times $t_{k}$ and $t_{l}$.
Note that even if $K(t_{k})=K(t_{l})$, state changes may still occur
between $t_{k}$ and $t_{l}$. Therefore, two realizations with the
same sequence $\left\{ K(t_{k})\right\} $ may still correspond to
different cases.

For the two-state problem in a RTP model, an example is to calculate
the second-order correlation function $\left\langle F\left(t_{2}\right)F\left(t_{1}\right)\right\rangle $.
There are totally six cases:

1)-4): $K\left(t_{2}\right)=K\left(t_{1}\right)=\mathrm{R}$ (or $\mathrm{T}$),
and there is (or no) state change between $t_{2}$ and $t_{1}$.

5): $K\left(t_{2}\right)=\mathrm{R}$, $K\left(t_{1}\right)=\mathrm{T}$
and 6): $K\left(t_{1}\right)=\mathrm{R}$, $K\left(t_{2}\right)=\mathrm{T}$.
These cases can be conveniently represented using diagrams. We place
vertices along a line, each corresponding to a time appearing in the
correlation function. The vertex is labeled according to the state
of the particle at the corresponding time, and a dash is used to connect
vertices when no state change occurs between the associated times.
Taking the second-order correlation function as an example, there
are only six cases that are required, 
\begin{equation}
\mathrm{R}_{t_{2}}\quad\mathrm{R}_{t_{1}},\hspace*{1.5em}\mathrm{R}_{t_{2}}-\mathrm{R}_{t_{1}},\hspace*{1.5em}\mathrm{T}_{t_{2}}\quad\mathrm{T}_{t_{1}},\hspace*{1.5em}\mathrm{T}_{t_{2}}-\mathrm{T}_{t_{1}}
\end{equation}
\begin{equation}
\mathrm{R}_{t_{2}}\quad\mathrm{T}_{t_{1}},\hspace*{1.5em}\mathrm{T}_{t_{2}}\quad\mathrm{R}_{t_{1}}
\end{equation}
in terms of the diagram expression.

In another example, one of the diagrams for the sixth-order correlation
function is given by 
\begin{equation}
\mathrm{R}_{t_{6}}-\mathrm{R}_{t_{5}}\quad\mathrm{T}_{t_{4}}-\mathrm{T}_{t_{3}}-\mathrm{T}_{t_{2}}-\mathrm{T}_{t_{1}}.\label{eq:sixth-order-diag-example}
\end{equation}
Here, this is the case where the particle remains in state $\mathrm{T}$
without any state change during the time from $t_{1}$ to $t_{4}$,
in state $\mathrm{R}$ without state changes during the time from
$t_{5}$ to $t_{6}$, whereas at least one (possibly more) state change
occurs sometime between $t_{4}$ and $t_{5}$.

This classification method enables a straightforward evaluation of
both the conditional expectation $\left\langle \cdot\mid\mathrm{diag}\right\rangle $
and the probability $\mathbb{P}(\mathrm{diag})$. Further details
are provided in Appendix \ref{sec:Calculation-of-the-Correlation-Function}.
In the end, the correlation function depends on the times $t_{k}$
only through an exponential dependence on time differences $e^{-c_{k}\left(t_{k+1}-t_{k}\right)}$
where $c_{k}$ can be calculated explicitly.

The presence of the harmonic potential introduces another exponential
dependence on time differences in Eq. \eqref{eq:MomentsGeneral}.
In particular, the factor $e^{-b\left(lt-\sum_{k}t_{k}\right)}$ can
be rewritten using the identity 
\begin{equation}
lt-\sum^{l}_{k=1}t_{k}=\sum^{l}_{k=1}k\left(t_{k+1}-t_{k}\right),\label{eq:integraltoconvolution}
\end{equation}
where $t_{l+1}=t$. This reformulation, together with the form of
the correlation function, allows the integral to be expressed in a
convolution form, which in turn enables a systematic evaluation of
its Laplace transform in time.

A recursion relationship can be established based on this method.
By cutting each diagram at the leftmost blank segment, the portion
to the right corresponds to a diagram already computed at lower order,
while the left segment forms a connected contribution that can be
evaluated explicitly. Define $L^{l}_{B}$ as the the Laplace transform
in time of contributions from all diagrams with $2l$ vertices and
ending with state $B$, excluding the prefactor $(2l)!/\left[\left(2lb+\xi\right)\xi\right]$.
The following recursive relation is concluded, 
\begin{equation}
L^{l}_{B}\left(\xi\right)=\sum^{l-1}_{k=0}\prod^{2l-1}_{m=2k+1}\frac{1}{\gamma_{B}+mb+\xi}v^{2\left(l-k\right)}_{B}M^{l-k}_{d}\sum_{B}g^{BA}_{k}L^{k}_{A}\left(\xi\right),\label{eq:l-recur}
\end{equation}
where the initial condition is\textcolor{red}{{} }$g^{BA}_{0}=\delta_{BA}$
and $L^{0}_{B}=\mathbb{P}\left(B\right)$. The factor $M^{l}_{d}$,
with $d$ denoting the spatial dimension, arises from the moments
of the active velocity and is given by 
\begin{equation}
M^{k}_{d}=\frac{\Gamma\left(1/2+k\right)\Gamma\left(d/2\right)}{\sqrt{\pi}\Gamma\left(d/2+k\right)}
\end{equation}
where $\Gamma$ is the gamma function. The factor $g$ represents
the contribution from the leftmost blank segment and is given by 
\begin{equation}
g^{BA}_{k}=\left(\left(\xi+2kb-\Lambda\right)^{-1}\right)_{BA}-\frac{\delta_{BA}}{2kb+\xi+\gamma_{A}},\label{eq:gk-ms}
\end{equation}
which follows from the Laplace transform in time of $P^{AB}$ defined
in eq. \ref{eq:P-Blank}.

This recursion allows the systematic computation of $L^{l}_{B}$ from
the initial conditions, after which the moments follow directly, 
\begin{equation}
\mathscr{L}\left(\left\langle x^{2l}\right\rangle \right)\left(\xi\right)=\frac{(2l)!}{\left(2lb+\xi\right)\xi}\sum_{B}L^{l}_{B}\left(\xi\right).
\end{equation}
For special cases such as the 1D theory presented below, all moments
can be calculated exactly, enabling the determination of the exact
characteristic function $\left\langle e^{iqx}\right\rangle $. By
inverse Fourier transform in $q$, the exact distribution function
$p\left(x\right)$ can then be obtained \citep{2025_ExactMomentsforaRunandTumbleParticlewithaFiniteTumbleTimeinaHarmonicTrap,2025_MomentAnalysisofTwoDimensionalActiveBrownianRunandTumbleParticles};
in other cases, approximations would need to be made. 

\subsection{Conditional moments of positional distribution }

\label{subsec:Conditional-moments}

Of central interest in this work is the conditional positional distribution
$p\left(x,t,A,v_{t}\right)$, given the conditions that $K\left(t\right)=A$
and $F\left(t\right)=v_{t}$. Namely, this is the positional distribution
at time $t$, of active particles that are in state $A$ with active
velocity $v_{t}$ at time $t$. In the RTP model, since the magnitudes
of the active velocity vector is fixed, only the direction of $v_{t}$
matters.

The first step is to find the conditional positional distribution
$p\left(x,t,A\right)$, given only the condition that $K\left(t\right)=A$,
integrated over $v_{t}$. Its moments are given by 
\begin{equation}
\left\langle x\left(t\right)^{l}|A\right\rangle =~l!\int_{0\leqslant t_{1}\leqslant...\leqslant t_{l}\leqslant t}\left(\prod^{l}_{i=1}dt_{i}\right)e^{-b\left(lt-\sum_{k}t_{k}\right)}\left\langle \prod^{l}_{k=1}F\left(t_{k}\right)|A\right\rangle .
\end{equation}
where $|A$ labels the condition $|K\left(t\right)=A$. A diagrammatic
expansion can be constructed, following a procedure similar to that
of the previous subsection, but conditional on $|A$. For the conditional
correlation function $\left\langle \prod_{k}F\left(t_{k}\right)|A\right\rangle $,
\begin{equation}
\left\langle \cdot|A\right\rangle =\sum_{\mathrm{diag}}\left\langle \cdot|\mathrm{diag},A\right\rangle \mathbb{P}\left(\mathrm{diag}|A\right)
\end{equation}
Again, in each case, $\mathrm{diag}$ is represented by a diagram
of vertices, labels, and segments. The only difference is that a case
now includes the state at time $t$. Diagrammatically, one additional
vertex, representing the state at time $t$, is appended to the left
of every diagram. Therefore, for example, a diagram for the sixth-order
conditional correlation function is: 
\begin{equation}
A_{t}-A_{t_{6}}-A_{t_{5}}\quad B_{t_{4}}-B_{t_{3}}-B_{t_{2}}-B_{t_{1}}
\end{equation}
which can be compared with \eqref{eq:sixth-order-diag-example}.

The detailed derivation is presented in Appendix \ref{subsec:Direction-Independent-Conditional-moments}.
In the end, many of the results for the unconditional moments in Sect.
\ref{subsec:The-Kac-method} can be reused. The Laplace transform
of the positional distribution in time is 
\begin{equation}
\mathscr{L}\left(\left\langle x^{2l}|A\right\rangle \right)\left(\xi\right)=\frac{(2l)!\sum_{B}h^{AB}_{2l}L^{l}_{B}\left(\xi\right)}{\xi},\label{eq:L-conditional}
\end{equation}
where $L^{l}_{B}$ is the same as in Eq. \ref{eq:l-recur}. In addition,
\begin{equation}
h^{AB}_{2l}=\frac{1}{\mathbb{P}\left(A\right)}\left[\left(\xi+2lb-\Lambda\right)^{-1}\right]_{AB},
\end{equation}
where the subscript $AB$ denotes the element of the inverted matrix
in row $B$ and column $A$. Note that the factor $1/(lb+\xi)$ is
no longer needed, as $h^{AB}_{2l}$ already accounts for the time
interval from $t_{2l}$ to $t$, whereas the factor $1/\xi$ is remains
essential.

As the next step, the direction-dependent conditional probability
distribution can now be determined, with the additional condition
$F(t)=v_{t}$, as required at the beginning of this section. For the
$\mathrm{T}$ state, the only possible value of $v_{t}$ is zero,
and therefore the condition $v_{t}=0$ is implicitly implied for the
T state.

Making use of the law of total expectation, 
\begin{equation}
\left\langle \cdot|A,v_{t}\right\rangle =\sum_{d_{B}}\left\langle \cdot|d_{B},A,v_{t}\right\rangle \mathbb{P}\left(d_{B}|A,v_{t}\right),
\end{equation}
where $d_{B}$ represents a diagram with $K(t_{l})=B$. The left-hand
side can be obtained. The derivation, along with the resulting diagrammatic
rules and recursive relations, is presented in Appendix \ref{subsec:Direction-Dependent-Conditional-Moments}.
It is worthwhile to highlight that:

1) $L^{l}_{B}$, defined earlier in Eq. \eqref{eq:l-recur}, can still
be used to recursively compute the direction-dependent conditional
moments.

2) Previously, the moments $\left\langle x(t)^{l}\right\rangle $
were non-vanishing only for even $l$, as a consequence of the left-right
symmetry of the projected problem. The condition $F(t)=v_{t}$ specifies
a direction for $v_{t}$, and thus the symmetry can no longer be exploited.
As a result, $\left\langle x(t)^{l}\mid A,v_{t}\right\rangle $ becomes
non-trivial for both even and odd $l$.

The main purpose of this work is to establish the significance of
the conditional moments and distributions, to find a method for calculating,
and to relate the results to the joint distribution in the Fokker--Planck
equation. Therefore, in the remainder of this work, we focus on the
1D case, for which analytic expressions can be obtained, in reference
to the Fokker-Planck equation. However, the methods developed so far
apply in arbitrary spatial dimensions.

\section{Run-and-tumble particles with zero tumble time}

\label{sec:Run-and-tumble-particle}

As an illustration of the approach presented above, we consider the
basic RTP model with zero tumble time. In this model, the particle
moves with a constant active velocity of magnitude $v$ for an exponentially
distributed time with rate $\gamma$, after which it randomly selects
a new direction for the active velocity. It then continues to run
for another exponentially distributed time with the same rate, and
this process is repeated thereafter. The model is ideal, as it yields
a simple exact result that illustrates the connection to the Fokker--Planck
equation.

The results presented in the previous sections are used here in the
special limit $\gamma_{T}\rightarrow\infty$. Only a single state
$\mathrm{R}$ is effectively present, with interruptions by short-lived
tumbles. For the entire process, the condition $K(t)=\mathrm{R}$
holds at all times with probability one. As a result, we may directly
work with the direction-dependent conditional distributions $\left\langle x^{2l}\mid v_{t}\right\rangle $.
For convenience, the label of state $\mathrm{R}$ is omitted in this
subsection, both in the diagrams and in the subscripts and superscripts
of $g$, $h$, $\gamma$, $v$, and $L$. In this case, $L$ takes
the form 
\begin{equation}
L^{l}(\xi)=v^{2l}\prod^{l-1}_{k=1}\frac{1}{2kb+\xi}\prod^{l}_{i=1}\frac{1}{\gamma+(2k-1)b+\xi}.
\end{equation}
Taking the limit, 
\begin{equation}
\lim_{t\rightarrow\infty}F(t)=\lim_{\xi\rightarrow0}\xi\mathscr{L}(F)(\xi),
\end{equation}
the exact characteristic function of positional distribution in the
steady state can then be obtained.

There are only two physical scales in the problem: one in time and
another in length. In the following, the unit system $b=1$ and $v=1$
is adopted. The symbols $+$ and $-$ are used to denote $v_{t}=v$
and $v_{t}=-v$, respectively. Expanding $e^{iqx}$ into a power series
in $x$ and exchanging the order of summation and expectation, the
characteristic function (i.e., the Fourier transform of positional
distribution in $x$) is obtained as: 
\begin{equation}
\left\langle e^{iqx}|\pm\right\rangle =\left(1\mp i\frac{d}{dq}\right)\,_{0}F_{1}\left(\frac{\gamma+1}{2};-\frac{q^{2}}{4}\right).
\end{equation}
By performing the inverse Fourier transform, the conditional steady-state
distributions 
\begin{equation}
p_{\pm}(x)=\frac{\Gamma\left(\frac{\gamma+1}{2}\right)}{\sqrt{\pi}\Gamma\left(\frac{\gamma}{2}\right)}(1\pm x)(1-x^{2})^{\gamma/2-1}.
\end{equation}
are finally obtained.

The results are now related to the Fokker--Planck equation. By differentiating
the solutions, one can verify that $f_{\pm}=p_{\pm}$ satisfy the
coupled Fokker--Planck equations 
\begin{equation}
0=\partial_{x}\left[\left(x-1\right)f_{+}\right]-\frac{\gamma}{2}f_{+}+\frac{\gamma}{2}f_{-},
\end{equation}
\begin{equation}
0=\partial_{x}\left[\left(x+1\right)f_{-}\right]-\frac{\gamma}{2}f_{-}+\frac{\gamma}{2}f_{+}.
\end{equation}
Therefore, for the 1D RTP problem, the conditional distribution is
indeed the solution of the Fokker--Planck equation.

Furthermore, this solution is obtained without assuming a priori that
a steady state exists. Instead, the steady state is obtained by taking
the limit $t\rightarrow\infty$. This provides clear evidence that
the Langevin system indeed approaches a steady state in an external
force field, providing a mathematical justification that is not obvious
at the outset.

Finally, we make a note about a seemingly trivial issue. For the construction
of the solution, the identification of $f_{\pm}=p_{\pm}$ made above
can be replaced by $f_{\pm}=Cp_{\pm}$ where $C$ is a constant. This
constant $C$ should be determined by the normalization of the total
probability distribution function $f=f_{+}+f_{-}$, $\int^{\infty}_{-\infty}f\left(x\right)dx=1$;
hence, $C=1/2$, as each of $p_{\pm}$ itself is already normalized.
This result, $f=(p_{+}+p_{-})/2$, can be re-composited as $f=\mathbb{P}(+)p_{+}+\mathbb{P}(-)p_{-}$,
as the left and right overall probabilities in the R state are equivalent,
$\mathbb{P}(\pm)=1/2$. In the nontrivial case where the probability
$\mathbb{P}$ for different states are not equivalent, such as the
case in the next section when the finite T state is involved, a careful
re-composition is essential for constructing the solution.

\section{1D run-and-tumble particle with finite tumble time}

\label{sec:1D-two-state}

As a final illustration, the two-state run-and-tumble model is revisited
in this section, with special emphasis on the 1D steady state where
the lifetimes of the states $\mathrm{R}$ and $\mathrm{T}$ are exponentially
distributed with rates $\gamma_{R}$ and $\gamma_{T}$, respectively.
The positional distribution was the focus of study in Ref. \citep{2025_ExactMomentsforaRunandTumbleParticlewithaFiniteTumbleTimeinaHarmonicTrap}.
Here, the exact conditional distributions for all states, $\mathrm{R}$,
$\mathrm{T}$, and $\pm$ (corresponding to the $\mathrm{R}$ state
with active velocities $\pm v$), are derived. This constitutes a
significant extension that provides further insight into the problem.

Although the system contains two states, $\mathrm{R}$ and $\mathrm{T}$,
we restrict attention to trajectories in which the particle is in
the $\mathrm{R}$ state at every time $t_{k}$ appearing in the correlation
function, since otherwise the correlation function vanishes. Consequently,
in the diagrammatic representation introduced in Section \ref{sec:Conditioned-moments},
all vertices, except for the leftmost one associated with time $t$,
must be labeled by $\mathrm{R}$. In this section, all $\mathrm{R}$
and $\mathrm{T}$ labels, including those appearing in superscripts
and subscripts, are omitted for brevity, except for $h^{RR}$ and
$h^{TR}$. In one dimension, the problem further simplifies since
$M^{k}_{1}=1$. Then, 
\begin{equation}
L^{l}\left(\xi\right)=\frac{v^{2}L^{l-1}\left(\xi\right)}{\gamma_{R}+(2l-1)b+\xi}\left[g_{l-1}+\frac{1}{\gamma_{R}+(2l-2)b+\xi}\right]
\end{equation}
where $g_{l}=g^{RR}_{l}$.

The exact characteristic function $\mathscr{L}\left(\left\langle e^{iqx}\right\rangle \right)$
can be evaluated in two cases. First, when $b=0$, the ratio $L^{l}(\xi)/L^{l-1}(\xi)$
becomes independent of $l$, and therefore $L^{l}(\xi)$ can be obtained
explicitly. The characteristic function $\left\langle e^{iqx}\right\rangle $
is then obtained, in units where $v=1$, to be: 
\begin{equation}
\mathscr{L}\left(\left\langle e^{iqx}|\pm\right\rangle \right)\left(\xi\right)=\frac{(\gamma_{R}+\gamma_{T}+\xi)(\gamma_{R}\pm iq+\xi)}{\gamma_{T}\left(\xi(\gamma_{R}+\xi)+q^{2}\right)+\xi\left((\gamma_{R}+\xi)^{2}+q^{2}\right)}\label{eq:charactristic-pp}
\end{equation}
\begin{equation}
\mathscr{L}\left(\left\langle e^{iqx}|T\right\rangle \right)\left(\xi\right)=\frac{\xi^{2}+(\gamma_{R}+\xi)(\gamma_{R}+\gamma_{T}+\xi)}{\gamma_{T}\left(\xi(\gamma_{R}+\xi)+q^{2}\right)+\xi\left((\gamma_{R}+\xi)^{2}+q^{2}\right)}\label{eq:characteristic-pt}
\end{equation}

Another case where the exact characteristic function can be obtained
is the steady state, where only $L(0)$ is needed. After performing
the inverse Fourier transform, the distribution can be obtained; the
explicit expressions are listed in Appendix \ref{sec:1DRTPSSD}, in
which the appearance of the singular peaks in the positional distribution
is carefully dissected.

From the exact solution, it is clear that 
\begin{equation}
p_{R}\left(x\right)=\frac{1}{2}\left(p_{+}\left(x\right)+p_{-}\left(x\right)\right)\label{ppm}
\end{equation}
Furthermore, although not straightforward to verify, the relation
\begin{equation}
p\left(x\right)=\mathbb{P}\left(\mathrm{R}\right)p_{R}\left(x\right)+\mathbb{P}\left(\mathrm{T}\right)p_{T}\left(x\right)\label{pdef}
\end{equation}
holds, where $\mathbb{P}(\mathrm{R})$ and $\mathbb{P}(\mathrm{T})$
denote $\mathbb{P}(K(t)=\mathrm{R})$ and $\mathbb{P}(K(t)=\mathrm{T})$,
respectively, for any $t>0$. These relations are consistent with
the law of total probability. Note $\mathbb{P}(\mathrm{T})\neq0$
for finite-tumble time RTPs.

\begin{figure*}
\centering\includegraphics[width=0.8\linewidth]{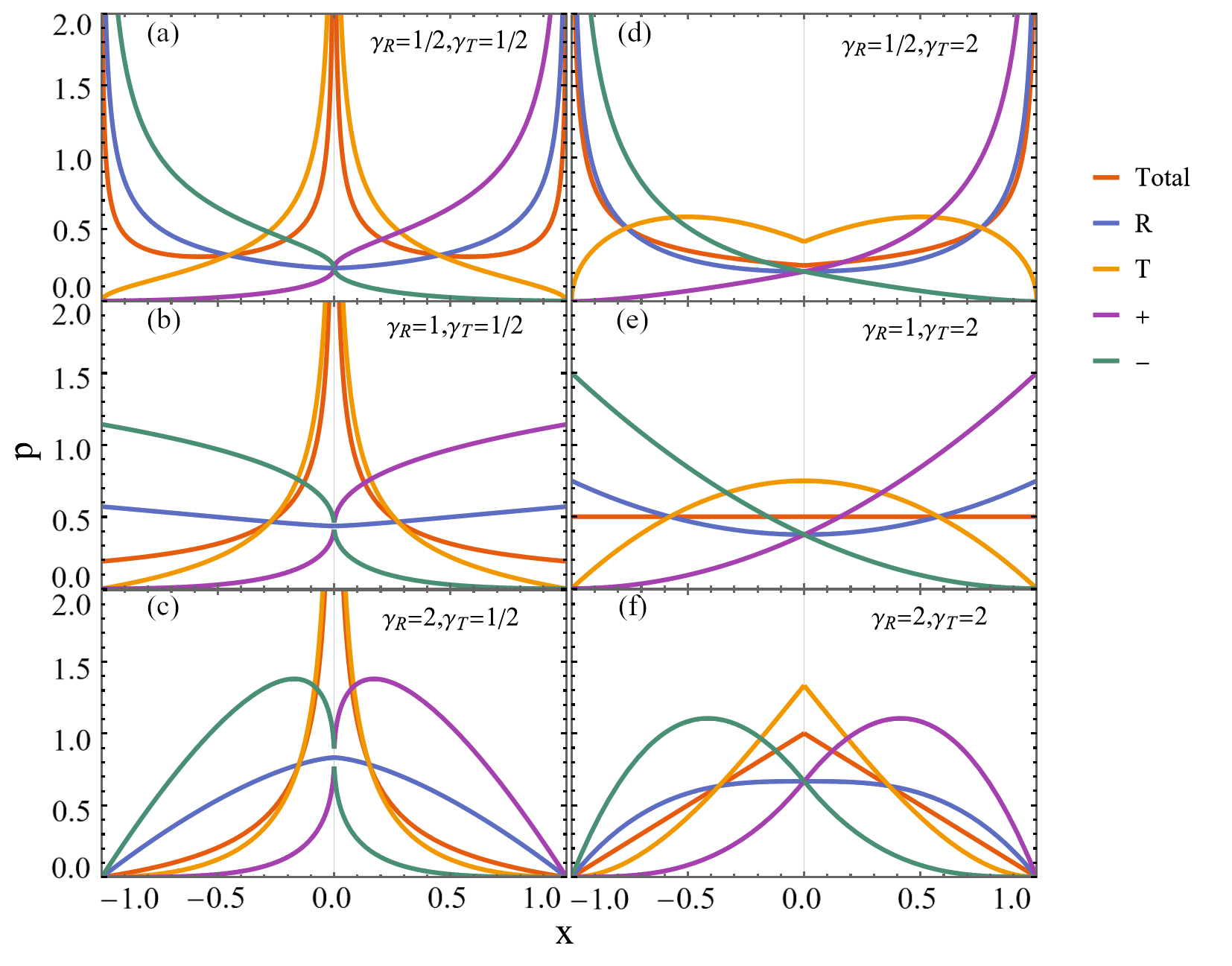} \caption{Steady-state distributions of the two-state RTP model at different
rates. Panels (a)--(c) illustrate cases with increasing $\gamma_{R}$
at a small $\gamma_{T}=1/2$, while panels (d)--(f) show cases with
increasing $\gamma_{R}$ at a larger $\gamma_{T}=2$. The total distribution
$p$, defined in \eqref{pdef}, is shown by the deep orange curves,
while the components $p_{R}$ and $p_{T}$ are shown in blue and orange,
respectively. The velocity-dependent distributions $p_{+}$ and $p_{-}$
appearing in \eqref{ppm} are shown in purple and green. For $\gamma_{T}<1$,
a singular peak at the center arises from the tumble state, accompanied
by a pronounced change in $p_{\pm}$ near the center. When $\gamma_{R}<1$,
the peaks at the boundaries originate from particles in the corresponding
run states. For $\gamma_{R}>1$, particles in the $\pm$ states are
more likely to be found midway between the center and the boundaries. }
\label{fig:2s1d} 
\end{figure*}

In a simple illustration of the conditional position distributions,
several examples with units $b=v=1$ are plotted in Fig. \ref{fig:2s1d}.
When $\gamma_{R}>1$, a particle in state $\mathrm{R}$ is more likely
to be found near the center. A particle in the $+$ state is most
likely to be found midway between $0$ and $1$.

Another feature is that, while $\gamma_{T}$ does not appear to affect
the $\mathrm{R}$ state significantly, it does influence the individual
$\pm$ states. For $\gamma_{T}<1$, a drastic change in the probability
density of both $\pm$ states is observed at the center, as shown
by the purple and green curves. In this regime, the derivative of
$p_{\pm}$ diverges, whereas their sum $p_{R}$ still has zero derivative.
The singularities in $p'_{\pm}$ cancel each other upon summation.

An interesting observation is when $\gamma_{T}$ is large and $\gamma_{R}$
is small, a particle in the $\mathrm{T}$ state is more likely to
be found midway between $0<x<1$ and $-1<x<0$. The explanation is
that, for small $\gamma_{R}$, the preceding $\mathrm{R}$ phase is
typically long, so that the particle is likely to be located near
a boundary when the tumble begins. Since the tumble state is typically
short-lived, the particle does not have sufficient time to relax toward
the center, and therefore remains located between the boundary and
the center. In this regime, $\mathbb{P}(\mathrm{T})$ is small, thus
this feature is almost invisible in the total positional distribution,
which instead exhibits two peaks near the boundaries and a shallow
minimum at the center.

The analytic solutions of $p_{\pm}(x)$ and $p_{T}(x)$ can then be
related to the the steady state solutions of the Fokker-Planck equation.
Three differential equations are written down, according the Fokker-Planck
theory \citep{2020_ExactStationaryStateofaRunandTumbleParticlewithThreeInternalStatesinaHarmonicTrap,2022_ExactPositionDistributionofaHarmonicallyConfinedRunandTumbleParticleinTwoDimensions},
\begin{equation}
\partial_{t}f_{+}=\partial_{x}\left(\left(bx-v\right)f_{+}\right)-\gamma_{R}f_{+}+\frac{\gamma_{T}}{2}f_{T}\label{eq:FPeq1}
\end{equation}
\begin{equation}
\partial_{t}f_{-}=\partial_{x}\left(\left(bx+v\right)f_{-}\right)-\gamma_{R}f_{-}+\frac{\gamma_{T}}{2}f_{T}
\end{equation}
\begin{equation}
\partial_{t}f_{T}=\partial_{x}\left(bxf_{T}\right)-\gamma_{T}f_{T}+\gamma_{R}\left(f_{+}+f_{-}\right)\label{eq:FPeq3}
\end{equation}
where $f_{\pm},f_{T}$ represent the density such that the total $f=f_{+}+f_{-}+f_{T}$
is normalized. In comparison with previous references, different rates,
$\gamma_{R}\neq\gamma_{T}$, and the time derivatives, are kept here
for the finite-time tumble problem.

The appearance that these coupled differential equations are linear
and first order does not, in general, lead to closed-form solutions.
When $b=0$, applying the Laplace transform in time and the Fourier
transform in space reduces the system to a linear algebraic form,
which is therefore exactly solvable. For $b>0$, the only known exact
solutions arise in special cases where the equations can be reduced
to simpler forms, as in \citep{2020_ExactStationaryStateofaRunandTumbleParticlewithThreeInternalStatesinaHarmonicTrap},
for $\gamma_{T}=\gamma_{R}$ or in the limit $\gamma_{T}\rightarrow\infty$.

The time-dependent functions $p_{\pm}(x)$ and $p_{T}(x)$, in their
Fourier and Laplace transformations Eq. \eqref{eq:charactristic-pp}
and Eq. \eqref{eq:characteristic-pt} or real-space expressions for
steady state in Appendix \ref{sec:1DRTPSSD}, are NOT the probability
functions appearing in the Fokker-Planck equations, in \eqref{eq:FPeq1}-\eqref{eq:FPeq3}.
The factors $\mathbb{P}(\pm)$ and $\mathbb{P}(\mathrm{T})$ are now
vital for constructing the solution, as $\mathbb{P}(\pm)$ are generally
not the same as $\mathbb{P}(\mathrm{T})$. It can be rigorously verified
by direct substitution that $f_{\pm}=\mathbb{P}(\pm)p_{\pm}$ for
the R state, $f_{T}=\mathbb{P}(\mathrm{T})p_{T}$ for the T state
satisfy the three above Fokker--Planck equations exactly.

\section{Summary and discussion}

\label{sec:Conclusion}

We analytically obtain the conditional moments of positional distribution
of RTPs with time dependence. The starting point is solving the Langevin
dynamics, following some of the basic ideas developed by Kac long
ago. The resulting conditional positional distributions are shown
to be equivalent to the joint distribution of position and velocity,
that satisfies the Fokker-Planck equation.

With this improvement in methodology, we not only rigorously confirm
our previous results in \citep{2025_ExactMomentsforaRunandTumbleParticlewithaFiniteTumbleTimeinaHarmonicTrap}
agrees with the Fokker-Planck equation, but also obtain more insights
of this problem. For the steady state of the 1D RTP model with finite
tumble time, as intuitively expected, the singular peaks in the distribution
function at $x=-1,0,1$, results from the particles in $-$, $\mathrm{T}$,
$+$ states, respectively. The derivative of the conditional distribution
$p_{\pm}$ diverges at $x=0$ for $\gamma_{T}<1$, causing a dramatic
change of conditional distribution density at the center. This interesting
features could not been studied with only the positional distribution.

While the physical model of active particles is the focus here, we
may also take out all the physical context, and view our treatment
from a purely mathematical angle. By finding a probability distribution
that satisfies a previously unsolved Fokker-Planck equation, we in
fact establish a probabilistic solution to a PDE system. While it
is common to study stochastic processes by solving the corresponding
Fokker-Planck equation, this opposite direction is much less studied
beyond Monte Carlo sampling. Furthermore, for the first time, the
importance of the conditional distribution is demonstrated.

In principle, the general method developed here is applicable to higher
spatial dimensions, where the moments of the conditional distribution
functions are systematically deduced. There are, on the other hand,
additional technical hurdles that need to be overcome in the diagram
expansion, when the position and active velocity are promoted into
vectors, which ultimately requires the construction of a vector-and-tensor
theory for an equation for equations such as \eqref{eq:MomentsGeneral}.
We therefore restrict ourselves to a complete 1D theory in this paper,
avoiding a discussion of 2D and 3D cases that would end inconclusively.

The method itself, which has the same mathematical starting point
as the Kac method for solving the Langevin equation, depends on the
linearity of the external force fields (harmonic potentials), which
has the strength of yielding analytic results. A recent study \citep{2025_RunandTumbleParticleinOneDimensionalPotentialsMeanFirstPassageTimeandApplications}
focuses on finding the mean first-passage time of a 1D RTP in an anharmonic
potential by tackling the Fokker--Planck equation. Furthering the
current analytic methodology to deal with the first-passage time problem
deserves future consideration.

\section*{Acknowledgments}

We are indebted to Dr. Rudolf Podgornik for his earlier participation
in the project. He has been a valued colleague and friend, and an
inspiration to us all. He will be sorely missed. This work was financially
supported by the National Key Research and Development Program of
China (2022YFA1405002), the National Natural Science Foundation of
China (Grant No. 12325405 \& T2221001), and the Zhejiang Key Laboratory
of Soft Matter Biomedical Materials (2025ZY01036 and 2025E10072).

\section*{Data Availability Statement}

Data sharing is not applicable to this article as no new data were
created or analyzed in this study.

\section*{AUTHOR DECLARATIONS}

\subsection*{Conflict of Interest}

The authors have no conflicts to disclose.

\subsection*{Author Contributions}

\textbf{Aoran Sun}: Conceptualization (lead); Investigation (lead);
Methodology (lead); Validation (equal); Writing -- original draft
(lead); Writing -- review \& editing (equal). \textbf{Fangfu Ye}:
Funding acquisition (lead); Project administration (lead); Resources
(lead); Supervision (equal); Validation (equal);Writing -- review
\& editing (equal). \textbf{Jeff Z. Y. Chen}: Validation (equal);
Supervision (equal); Writing -- review \& editing (equal). 
\begin{widetext}

\appendix

\section{Calculation of the Correlation Function }

\label{sec:Calculation-of-the-Correlation-Function}

In Sect.~\ref{subsec:The-Kac-method}, it is stated that the case-by-case
discussion allows for a straightforward calculation of the correlation
function $\left\langle \prod_{k}F\left(t_{k}\right)\right\rangle $.
Details are provided here.

For every case, both the conditional expectation $\left\langle \prod_{k}F\left(t_{k}\right)\mid\mathrm{diag}\right\rangle $
and the probability $\mathbb{P}\left(\mathrm{diag}\right)$ are needed.
We start with the conditional expectation $\left\langle \prod_{k}F\left(t_{k}\right)\mid\mathrm{diag}\right\rangle $.
As $F$ is not correlated between segments, the expectation is broken
down to the product of expectation on each segments, on which $F$
is the same. For the RTP model, a segment passing through $2k$ vertices
in state $A$ contributes $v^{2k}_{A}M^{k}_{d}$ where $d$ is the
spatial dimension and 
\begin{equation}
M^{k}_{d}=\frac{\Gamma\left(1/2+k\right)\Gamma\left(d/2\right)}{\sqrt{\pi}\Gamma\left(d/2+k\right)}.\label{eq:MKD}
\end{equation}
the $\Gamma$ is the gamma function. Segments passing through an odd
number of vertices contribute a factor of zero to the contribution
of the diagram, due to the left-right symmetry of the projected problem
in the choice of $F$ at the beginning of the R state. One example
is the diagram: 
\begin{equation}
A-A\quad B-B-B-B
\end{equation}
for which the conditional expectation is: 
\begin{equation}
\left\langle \prod^{6}_{k=1}F\left(t_{k}\right)|A-A\quad B-B-B-B\right\rangle =v^{2}_{A}v^{4}_{B}M^{1}_{d}M^{2}_{d}.
\end{equation}
Note that subscripts indicating time have been omitted, and we follow
the convention that later times are placed to the left.

For the probability $\mathbb{P}(\mathrm{diag})$, the diagram is partitioned
into individual events. Each event accounts the presence or absence
of state changes between $t_{k}$ and $t_{k+1}$ for all $k$. Afterward
the Markov property is applied, to multiply the probabilities of all
these individual events together, obtaining the probability of this
case $\mathbb{P}(\mathrm{diag})$. If a diagram starts in state $A$,
the factor $\mathbb{P}(A)$ is also included. Moreover, if $t_{k+1}$
is connected to $t_{k}$ with label $A$, the corresponding survival
probability $e^{-\gamma_{A}(t_{k+1}-t_{k})}$ is included as a multiplicative
factor. Otherwise, if the vertex $A$ at $t_{k+1}$ is not connected
to the vertex $B$ at time $t_{k}$, multiply 
\begin{align}
 & P^{AB}\left(t_{k+1}-t_{k}\right)\nonumber \\
= & \mathbb{P}\left(K\left(t_{k+1}\right)=A|K\left(t_{k}\right)=B\right)-\delta_{AB}e^{-\gamma_{A}\left(t_{k+1}-t_{k}\right)}.\label{eq:P-Blank}
\end{align}
The $\delta_{AB}$ term is necessary due to the following consideration.
In the case $A=B$, care must be taken to exclude the probability
that the particle does not change states over the entire time interval;
otherwise, the corresponding segments would remain connected. On the
other hand, when $A\neq B$, $K\left(t_{k+1}\right)=A$ and $K\left(t_{k}\right)=B$
naturally implies state changes, and no additional term is required.
To finish calculating $P$ in Eq. \eqref{eq:P-Blank}, we also need:
\begin{equation}
\mathbb{P}\left(K\left(t_{k+1}\right)=A|K\left(t_{k}\right)=B\right)=\left(\exp\left[\Lambda\left(t_{k+1}-t_{k}\right)\right]\right)_{AB},
\end{equation}
in which a matrix exponential function has been used, where $\Lambda$
are given by $\Lambda_{AA}=-\gamma_{A}$ and $\Lambda_{AB}=\gamma_{AB}$
for $B\neq A$, as discussed in Sect. \eqref{subsec:Langevin-equation}.
This conditional probability follows by solving the differential equations,
\begin{equation}
\dot{p}_{X}\left(t\right)=-\gamma_{X}p_{X}\left(t\right)+\sum_{Y\neq X}\gamma_{XY}p_{Y}\left(t\right),\label{eq:P-ode}
\end{equation}
subject to the initial condition $p_{X}(t_{k})=\delta_{BX}$, and
subsequently evaluating the conditional probability $p_{A}\left(t_{k+1}\right)$.

In the example diagram above, the probability is: 
\begin{align}
 & \mathbb{P}\left(A-A\quad B-B-B-B\right)\nonumber \\
= & \mathbb{P}\left(B\right)e^{-\gamma_{A}\left(t_{6}-t_{5}\right)}e^{-\gamma_{B}\left(t_{4}-t_{1}\right)}P^{AB}\left(t_{5}-t_{4}\right)
\end{align}
Combining the probability and conditional expectation, the contribution
of this diagram to the correlation function is: 
\begin{equation}
v^{2}_{A}v^{4}_{B}M^{1}_{d}M^{2}_{d}\mathbb{P}\left(B\right)e^{-\gamma_{A}\left(t_{6}-t_{5}\right)}e^{-\gamma_{B}\left(t_{4}-t_{1}\right)}P^{AB}\left(t_{5}-t_{4}\right)
\end{equation}
To complete the calculation of the correlation function, the contributions
from all possible diagrams need to be summed.

\section{Calculation of the Conditional Moments }

\label{sec:Calculation-for-Conditional-Moments}

In Subsection \ref{subsec:Conditional-moments}, it is stated that
the conditional moments can be evaluated on a case-by-case basis,
in analogy with the corresponding unconditional problem. The procedure
involves two steps: first, conditional on $K(t)=A$ without specifying
$F(t)$, and then, specializing to $F(t)=v_{t}$.

\subsection{Conditional Moments without specifying $F(t)$ }

\label{subsec:Direction-Independent-Conditional-moments}

This subsection is focused on the calculation of the conditional moments.
\begin{equation}
\left\langle x\left(t\right)^{l}|A\right\rangle =~l!\int_{0\leqslant t_{1}\leqslant...\leqslant t_{l}\leqslant t}\left(\prod^{l}_{i=1}dt_{i}\right)e^{-b\left(lt-\sum_{k}t_{k}\right)}\left\langle \prod_{k}F\left(t_{k}\right)|A\right\rangle .
\end{equation}
where $|A$ stands for the condition $|K\left(t\right)=A$. The idea
is still a case by case discussion 
\begin{equation}
\left\langle \cdot|A\right\rangle =\sum_{\mathrm{diag}}\left\langle \cdot|\mathrm{diag},A\right\rangle \mathbb{P}\left(\mathrm{diag}|A\right)
\end{equation}
and as mentioned in Sect. \ref{subsec:Conditional-moments}, now $\mathrm{diag}$
includes the state at time $t$.

For any given diagram, the conditional expectation $\left\langle \cdot|\mathrm{diag},A\right\rangle $
can still be computed analytically by multiplying the contributions
of individual segments. A segment labeled $A$ that passes through
$2k$ vertices contributes $v^{2k}_{A}M^{k}_{d}$. Note that events
occurring at the final time $t$ are included in $\mathrm{diag}$,
but need not be included when evaluating the conditional expectation.

For the conditional probability $\mathbb{P}\left(\mathrm{diag}|A\right)$,
the Bayes formula: 
\begin{equation}
\mathbb{P}\left(\mathrm{diag}|A\right)\mathbb{P}\left(A\right)=\mathbb{P}\left(A|\mathrm{diag}\right)\mathbb{P}\left(\mathrm{diag}\right)
\end{equation}
is utilized. In any diagram, conditioned on the state at time $t$
being $A$, $\mathbb{P}\left(A|\mathrm{diag}\right)=1$. Since the
Markov chain is initialized in equilibrium, $\mathbb{P}(A)$ is time-independent.
The probability $\mathbb{P}(\mathrm{diag})$ is evaluated as in the
previous section, with the difference that an additional vertex is
present, accounting for events between $t_{\ell}$ and $t$, and hence
one extra event contribution, either a survival probability $e^{-\gamma_{A}(t-t_{l})}$
when connected, or a $P^{AB}$ when disconnected, must be included.

Based on these results, a diagrammatic law can be formulated, analogous
to the construction in the last Appendix. The expression for the unconditional
correlation function obtained in the previous section can be reused.
Let $d_{B}$ denote the set of diagrams for which $K(t_{\ell})=B$.
For $A\neq B$, the leftmost vertex at time $t$ must be disconnected
from the remainder of the diagram in $d_{B}$. Consequently, using
the result for $\mathbb{P}(\mathrm{diag})$, the additional vertex
contributes a multiplicative factor $\mathbb{P}\left(A|K\left(t_{l}\right)=B\right)/\mathbb{P}\left(A\right)$,
which can be calculated the same way by solving the differential equations
in \eqref{eq:P-ode}.

On the other hand, if $A=B$, then each diagram in which $t$ is connected
to $t_{\ell}$ has a unique counterpart in which $t$ is disconnected
from $t_{\ell}$, with all other features identical. Since the conditional
expectations of these paired diagrams coincide, their contributions
may be combined through summation of probabilities. Consequently,
the additional condition produces an overall multiplicative factor
$\mathbb{P}\left(A|K\left(t_{l}\right)=A\right)/\mathbb{P}\left(A\right)$
to the contribution, compared with unconditional problems.

Summarizing the contributions, relative to the unconditional problem,
a diagram with $K(t_{\ell})=B$ acquires an additional multiplicative
factor in the conditioned correlation function for $K(t)=A$, given
by 
\begin{equation}
H^{AB}_{k}=\frac{\left(\exp\left(\Lambda\left(t_{k+1}-t_{k}\right)\right)\right)_{AB}}{\mathbb{P}\left(A\right)},
\end{equation}
with the convention $t_{\ell+1}=t$. This factor depends only on the
intermediate state $K(N(t_{\ell}))$. By distributive property of
multiplication over summation, one may first compute the unconditioned
diagrams without the appended vertex at $t$ as usual, and subsequently
multiply by the factor $H$.

After performing the Laplace transform in time, this additional factor
is mapped to 
\begin{equation}
h^{AB}_{2l}=\frac{1}{\mathbb{P}\left(A\right)}\left(\left(\xi+2lb-\Lambda\right)^{-1}\right)_{AB}.
\end{equation}
Finally, we obtain 
\begin{equation}
\mathscr{L}\left(\left\langle x^{2l}|A\right\rangle \right)\left(\xi\right)=\frac{(2l)!\sum_{Y}h^{AB}_{2l}L^{l}_{B}\left(\xi\right)}{\xi},
\end{equation}
which is Eq. \ref{eq:L-conditional} in the main text.

\subsection{Conditional Moments with $F(t)=v_{t}$ specified}

\label{subsec:Direction-Dependent-Conditional-Moments}

In comparison with the above, an additional condition $F(t)=v_{t}$
is introduced. The procedure remains analogous, with the law of total
expectation now taken conditional on $|A,v_{t}$, 
\begin{equation}
\left\langle \cdot|A,v_{t}\right\rangle =\sum_{d_{B}}\left\langle \cdot|d_{B},A,v_{t}\right\rangle \mathbb{P}\left(d_{B}|A,v_{t}\right),
\end{equation}
where $d_{B}$ represents a diagram with $K\left(t_{l}\right)=B$.

For the conditional expectation, $\left\langle \prod_{k}F\left(t_{k}\right)|d_{B},A,v_{t}\right\rangle $,
if $t$ is not connected to $t_{\ell}$ (i.e., if the leftmost segment
does not pass through $t$), then the additional condition $F(t)=v_{t}$
has no effect on events prior to $t_{\ell}$. For such a diagram $d_{B}$,
one therefore has $\left\langle \prod_{k}F\left(t_{k}\right)|d_{B},A,v_{t}\right\rangle =\left\langle \prod_{k}F\left(t_{k}\right)|d_{B},A\right\rangle $.
Otherwise, the calculation proceeds in a manner very similar to the
unconditional case. The only difference arises with the leftmost segment:
if it passes through the vertex at time $t$ and additional $k$ vertices:
it contributes a factor $v^{k}_{t}$ instead of $M^{k/2}_{d}$, since
the active velocity is now fixed by condition, rather than randomly
sampled. As a result, diagrams in which the leftmost segment passes
through $t$ and an odd number of additional vertices yield non-zero
contributions. In contrast, segments not involving $t$ that pass
through an odd number of vertices still contribute zero. Consequently,
odd moments of the direction-dependent conditional distributions can
be non-zero.

For the probability $\mathbb{P}\left(d_{B}|A,v_{t}\right)$, the condition
$v_{t}$ does not affect the probability, $\mathbb{P}\left(d_{B}|A,v_{t}\right)=\mathbb{P}\left(d_{B}|A\right)$,
and thus it is known. Therefore, after performing the Laplace transform
in time, we can summarize these steps into the diagrammatic law below:
To compute the conditional $l$-th order moments with final state
$A$ and (projection of) active velocity $v_{t}$, following steps
are taken:

1) Draw $l+1$ vertices on a line and connect vertex $2k$ to vertex
$2k-1$ for all $k\leq\left(l+1\right)/2$;

2) Connect vertex $2k$ to vertex $2k+1$ for some arbitrary chosen
$k$ (with $k\leq l/2$);

3) Assign a state label to each connected segment;

4) For each solid segment connecting vertex $j$ to vertex $k$ with
label $A$, assign the factor 
\begin{equation}
v^{\left|k-j\right|+1}_{A}\prod^{k-1}_{n=j}\frac{1}{\gamma_{A}+bn+\xi}
\end{equation}
where $\xi$ is the Laplace variable, $\gamma_{A}$ is the transition
rate of state $A$, and $v_{A}$ is the magnitude of active velocity
in state $A$. In addition, for the leftmost segment passing through
the vertex at time $t$ and an additional $k$ vertices, multiply
by $v^{k}_{t}$; for all other segments passing through $2k$ vertices,
multiply by $M^{k}_{d}$ as in Eq.~\ref{eq:MKD}. For every blank
interval between vertex $2k$ and vertex $2k+1$ with label $B$ on
the left and $A$ on the right, include a factor $g^{BA}_{k}$ as
defined in Eq.~\ref{eq:gk-ms}.

5) Multiply an additional factor $\frac{l!}{\mathbb{P}(A)\xi}$ to
obtain the contribution from this diagram.

6) Sum over all configurations in 2) and 3) to obtain the direction
dependent conditional $l$ moment.

For the recursive relation, $L^{k}_{A}(\xi)$ as defined in Eq.~\ref{eq:l-recur}
can be reused. To recover the moments from $L$, the following facts
are noted. If $A\neq B$, then $t$ cannot be connected to $t_{\ell}$,
and the additional condition $v_{t}$ has no effect. If $A=B$, then
$t$ may or may not be connected to $t_{\ell}$. When they are not
connected, the calculation proceeds in the same manner as in the case
$A\neq B$. When they are connected, however, the contribution to
the conditional expectation from the leftmost segment is modified,
and the diagram must again be decomposed from the leftmost blank.
Therefore, for even moments: 
\begin{align}
\mathscr{L}\left(\left\langle x^{2l}|A,v_{t}\right\rangle \right)\left(\xi\right)= & \frac{\left(2l\right)!}{\mathbb{P}\left(A\right)\xi}\sum^{l}_{k=0}\left(\prod^{2l}_{m=2k+1}\frac{1}{\gamma_{A}+mb+\xi}\right)\sum_{X}\left(g^{AX}_{k}L^{k}_{X}\left(\xi\right)\right)v^{2l-2k}_{t}\nonumber \\
= & \frac{\left(2l\right)!\sum_{X}g^{AX}_{l}L^{l}_{X}\left(\xi\right)}{\mathbb{P}\left(A\right)\xi}+\frac{2l\left(2l-1\right)v^{2}_{t}\mathscr{L}\left(\left\langle x^{2l-2}|A,v_{t}\right\rangle \right)\left(\xi\right)}{\left(\gamma_{A}+2lb+\xi\right)\left(\gamma_{A}+\left(2l-1\right)b+\xi\right)},
\end{align}
and for the odd moments: 
\begin{align}
\mathscr{L}\left(\left\langle x^{2l+1}|A,v_{t}\right\rangle \right)\left(\xi\right)= & \frac{\left(2l+1\right)!}{\mathbb{P}\left(A\right)\xi}\sum^{l}_{k=0}\left(\prod^{2l+1}_{m=2k+1}\frac{1}{\gamma_{A}+mb+\xi}\right)\sum_{X}\left(g^{AX}_{k}L^{k}_{X}\left(\xi\right)\right)v^{2l+1-2k}_{t}\nonumber \\
= & \frac{\left(2l+1\right)!v_{t}\mathscr{L}\left(\left\langle x^{2l}|A,v_{t}\right\rangle \right)\left(\xi\right)}{\gamma_{A}+\left(2l+1\right)b+\xi}.
\end{align}
Specifically, in one dimension, $M^{l-k}_{1}=1$ and $v^{2l-2k}_{t}=v^{2l-2k}_{A}$,
$v^{2l+1-2k}_{t}=\pm v^{2l+1-2k}_{A}$, therefore these relation can
be simplified into: 
\begin{align}
\mathscr{L}\left(\left\langle x^{2l}|A,\pm v_{A}\right\rangle \right)\left(\xi\right)= & \frac{\left(2l\right)!\sum_{X}h^{AX}_{2l}L^{l}_{X}\left(\xi\right)}{\xi},
\end{align}
\begin{align}
\mathscr{L}\left(\left\langle x^{2l+1}|A,\pm v_{A}\right\rangle \right)\left(\xi\right)= & \pm\frac{\left(2l+1\right)!v_{A}\sum_{X}\left(h^{AX}_{2l}L^{l}_{X}\left(\xi\right)\right)}{\xi\left(\gamma_{A}+(2l+1)b+\xi\right)}.
\end{align}

\section{Steady-State Conditional Positional Distribution for 1D RTP with
Finite Tumble Time in a Harmonic Potential}

\label{sec:1DRTPSSD}

In Section \ref{sec:1D-two-state}, the exact steady-state direction-dependent
conditional positional distributions for a one-dimensional RTP with
finite tumble-time in a harmonic potential are obtained. The explicit
expressions for the R state distribution are, 
\begin{equation}
p_{\pm}\left(x\right)=p_{R}\left(x\right)\pm x\frac{\gamma_{R}+\gamma_{T}}{\gamma_{T}}p\left(x\right)
\end{equation}
where, 
\begin{equation}
p_{R}\left(x\right)=\frac{\sqrt{\pi}\left(\gamma_{R}+\gamma_{T}\right)\Gamma\left(\frac{\gamma_{R}+\gamma_{T}}{2}\right)\Gamma\left(\frac{1+\gamma_{R}}{2}\right)}{\gamma_{T}\Gamma\left(\frac{\gamma_{T}}{2}\right)\Gamma\left(\frac{\gamma_{R}}{2}\right)\cos\frac{\pi\gamma_{T}}{2}}\left\{ \frac{_{2}F_{1}\left(\frac{2-\gamma_{R}}{2},\frac{1-\gamma_{R}-\gamma_{T}}{2};\frac{1-\gamma_{T}}{2};x^{2}\right)}{\Gamma\left(\frac{1+\gamma_{R}+\gamma_{T}}{2}\right)\Gamma\left(\frac{1-\gamma_{T}}{2}\right)}-\frac{\left|x\right|^{\gamma_{T}+1}{}_{2}F_{1}\left(\frac{2-\gamma_{R}}{2},\frac{3-\gamma_{R}+\gamma_{T}}{2};\frac{3+\gamma_{T}}{2};x^{2}\right)}{\Gamma\left(\frac{\gamma_{R}-\gamma_{T}-1}{2}\right)\Gamma\left(\frac{3+\gamma_{T}}{2}\right)}\right\} ,
\end{equation}
and: 
\begin{equation}
p\left(x\right)=\frac{\sqrt{\pi}\Gamma\left(\frac{\gamma_{R}+\gamma_{T}}{2}\right)\Gamma\left(\frac{1+\gamma_{R}}{2}\right)}{\Gamma\left(\frac{\gamma_{T}}{2}\right)\Gamma\left(\frac{\gamma_{R}}{2}\right)\cos\frac{\pi\gamma_{T}}{2}}\left\{ \frac{\left|x\right|^{\gamma_{T}-1}{}_{2}F_{1}\left(\frac{2-\gamma_{R}}{2},\frac{1-\gamma_{R}+\gamma_{T}}{2};\frac{1+\gamma_{T}}{2};x^{2}\right)}{\Gamma\left(\frac{1+\gamma_{R}-\gamma_{T}}{2}\right)\Gamma\left(\frac{1+\gamma_{T}}{2}\right)}-\frac{_{2}F_{1}\left(\frac{2-\gamma_{R}}{2},\frac{3-\gamma_{R}-\gamma_{T}}{2};\frac{3-\gamma_{T}}{2};x^{2}\right)}{\Gamma\left(\frac{\gamma_{R}+\gamma_{T}-1}{2}\right)\Gamma\left(\frac{3-\gamma_{T}}{2}\right)}\right\} .
\end{equation}
The function $p_{R}\left(x\right)$ is the positional distribution
in the R state, and $p$ is the unconditional positional distribution.
For the T state, 
\begin{equation}
p_{T}\left(x\right)=\frac{\sqrt{\pi}\left(\gamma_{R}+\gamma_{T}\right)\Gamma\left(\frac{\gamma_{R}+\gamma_{T}}{2}\right)\Gamma\left(\frac{1+\gamma_{R}}{2}\right)}{2\Gamma\left(\frac{\gamma_{T}}{2}\right)\Gamma\left(\frac{\gamma_{R}}{2}\right)\cos\frac{\pi\gamma_{T}}{2}}\left\{ \frac{2\left|x\right|^{\gamma_{T}-1}{}_{2}F_{1}\left(-\frac{\gamma_{R}}{2},\frac{1-\gamma_{R}+\gamma_{T}}{2};\frac{1+\gamma_{T}}{2};x^{2}\right)}{\gamma_{R}\Gamma\left(\frac{1+\gamma_{R}-\gamma_{T}}{2}\right)\Gamma\left(\frac{1+\gamma_{T}}{2}\right)}-\frac{_{2}F_{1}\left(\frac{2-\gamma_{R}}{2},\frac{1-\gamma_{R}-\gamma_{T}}{2};\frac{3-\gamma_{T}}{2};x^{2}\right)}{\Gamma\left(\frac{\gamma_{R}+\gamma_{T}+1}{2}\right)\Gamma\left(\frac{3-\gamma_{T}}{2}\right)}\right\} .
\end{equation}
The parameters $\gamma_{T}$ and $\gamma_{R}$ are the transition
rates of the tumble and run states, respectively. The function $_{2}F_{1}$
is the hypergeometric function \citep{2009_CRCEncyclopediaofMathematics}.

These functions are plotted in Fig. \ref{fig:2s1d} in the text. Mathematical
singularities arise in the exact solutions. Since $_{2}F_{1}\left(a,b;c;0\right)=1$,
$p_{R}(0)$ and $p_{\pm}(0)$ remain finite for all parameter values.
Only the quantities $p(0)$ and $p_{T}(0)$ diverge for $\gamma_{T}<1$
due to the factor $|x|^{\gamma_{T}-1}$. This is consistent with the
physical intuition that the tumble state is responsible for the singular
behavior at the origin. For sufficiently large $\gamma_{R}$, the
run-state distribution may exhibit a smooth peak near the center,
but it does not develop any true singularity.

At the boundary $x=1$, a sufficient condition for the finiteness
of $_{2}F_{1}\left(a,b;c;1\right)$ is $c>a+b$. For $p_{T}$, this
condition reduces to $\gamma_{R}>0$, which is always satisfied. For
$p_{R}$, $p_{\pm}$, and $p$, it becomes $\gamma_{R}>1$. In this
case, using the identity 
\begin{equation}
_{2}F_{1}(a,b;c;1)=\frac{\Gamma(c)\Gamma(c-a-b)}{\Gamma(c-a)\Gamma(c-b)},
\end{equation}
one can show that all densities vanish at the boundary: 
\begin{equation}
p_{\pm}\left(\pm1\right)=p_{R}\left(\pm1\right)=p_{T}\left(\pm1\right)=0
\end{equation}

For $\gamma_{R}<1$, the singular behavior near the boundary arises
entirely from the run state. In particular, $p_{+}(1)$ diverges and
provides the leading contribution to the singularity at the right
boundary $x=1$, whereas for $p_{-}(1)$ the singular terms cancel,
yielding $p_{-}(1)=0$. This is consistent with the physical intuition
that only particles in the $+$ state can accumulate near the right
boundary. 
\end{widetext}

\bibliography{lib}

\end{document}